\documentclass[12pt]{iopart}

\usepackage{iopams}
\usepackage{epsfig}

\def\brho{{\boldsymbol\rho}}
\def\bfg{{\boldsymbol g}}
\def\bfk{{\boldsymbol k}}
\begin{document}

\title[Manipulation of coherent atom waves using accelerated 2D optical lattices]
{Manipulation of coherent atom waves using accelerated two-dimensional optical lattices}

\author{Wei-Chih Ting, Dian-Jiun Han, and Shin-Tza Wu}
\address{Department of Physics, National Chung Cheng University,
Chiayi 621, Taiwan}

\ead{phystw@gmail.com}

\begin{abstract}
We study the dynamics of Bose-Einstein condensates in accelerated
two-dimensional optical square lattices by numerically solving the
Gross-Pitaevskii equation. We consider the regime with negligible
mean-field interactions and examine in detail the pulses of atom
clouds ejected from the condensate due to Landau-Zener tunnelling.
The pulses exhibit patterned structures that can be understood from
the momentum-space dynamics of the condensate. Aside from conceiving
realization of a pulsed two-dimensional atom laser, we demonstrate
that, by exploring the band structure of the lattice, Landau-Zener
tunnelling and Bragg reflection of the condensate inside the optical
lattice can provide means for manipulation of coherent atom waves.
\end{abstract}

%\pacs{03.75.Be, 03.75,Kk, 03.75.Lm, 03.75.Pp, 37.25.+k, 67.85.De, 67.85.Hj, }
%\submitto{\NJP}

%\tableofcontents

\maketitle

\section{Introduction}
%=====================
The past two decades have witnessed the fruitful interplay between
condensed matter physics and ultracold atom physics. While cold atom
systems have provided new arena for condensed matter physics, at the
same time, concepts and ideas from condensed matter physics have
opened up new directions for cold atom physics. For instance, by
loading ultracold atoms into optical potentials due to standing
laser waves (i.e.~optical lattices), it is possible not only to
build up ``quantum simulators" for condensed matter systems (see,
for example, \cite{bloch}), but also to control and manipulate cold
atoms utilizing concepts from traditional solid-state systems (see,
for instance, \cite{MO}). In this work, we wish to further explore
the latter lines of investigation.

Electrons in solids are subject to periodic potentials due to the
array of ion cores that constitutes the lattice. In ideal
situations, the electron eigenfunctions take the form of a
particular structure that is called the Bloch wave, and the
corresponding eigen-energies cluster into energy bands that are
separated by energy gaps (see e.g.~\cite{AM}). As a consequence,
electrons in solids can behave quite differently compared with those
in free space. For instance, when a weak uniform static electric
field is applied to the solid, instead of being uniformly
accelerated, the electrons would carry out oscillatory motions (the
Bloch oscillation; see e.g.~\cite{Ziman}). This oscillatory motion
can be understood as Bragg reflections of the Bloch states in
momentum space (or $\bfk$-space) at regions where energy gaps occur.
If the bias field is large, the electrons can gain sufficient energy
that would lead to finite probability for tunnelling across the
energy gaps, resulting in transitions to higher energy bands (the
Landau-Zener tunnelling; see, for example, \cite{Ziman}). In this
work, we wish to consider the cold atom counterpart of the above
construction. Namely, we shall consider a cloud of Bose-Einstein
condensate subject to a periodic potential due to standing laser
waves that form an optical lattice. We shall examine dynamics of the
condensate under the action of an external bias that corresponds to
the uniform static electric field above. Experimentally this bias
can be furnished in a number of ways: one could make use of direct
gravity pull on the atoms, or introduce time-dependent frequency
shifts between the lattice beams so that an effective acceleration
on the atoms occurs (see e.g.~\cite{RSQ,PS}). For spin polarized
condensates, it is also possible to generate the acceleration by
means of magnetic field gradients \cite{Couvert}.

The dynamics of ultracold atoms in one-dimensional (1D) optical
lattices subject to uniform accelerations have been investigated
extensively \cite{MO}. In addition to the observation of Bloch
oscillations \cite{Dahan}, Wannier-Stark ladders \cite{WS}, and
Landau-Zener tunnelling \cite{LZ1d} for condensates in accelerated
1D optical lattices, possible realizations for 1D ``atom laser" have
also been achieved \cite{AK} using such systems.\footnote{For other
realizations of atom lasers (involving and not involving accelerated
optical lattices), see e.g.~\cite{Couvert} and references therein.}
By loading Bose condensed $^{87}$Rb atoms into a vertical 1D optical
lattice, Anderson and Kasevich have demonstrated that for
appropriate lattice strengths, pulses of atom clouds can tunnel out
of the optical lattice due to the gravity pull. The generation of
pulses of coherent atom waves thus suggests the system being an 1D
pulsed atom laser \cite{AK,AK2}. For two-dimensional (2D) systems,
there have also been studies of Bloch oscillations, Landau-Zener
tunnelling (at lower energy bands; cf.~below) \cite{KK,germany04},
and Bloch-Zener oscillations \cite{germany07} of ultracold atoms in
accelerated 2D optical lattices.\footnote{There have also been
related theoretical and experimental studies in photonic crystals
\cite{Trompeter,Shchesnovich,SDK,Dreisow}.} The richer dynamics of
the condensates in 2D lattices has been proposed to be possible
means for manipulation of the coherent atom waves
\cite{KK,germany04,germany07}. Along these lines of investigation,
we shall examine in this work the Landau-Zener tunnelling of atom
waves within 2D optical lattices. More specifically, we will be
interested in physical configurations similar to the 1D system
considered by Anderson and Kasevich \cite{AK,AK2}. To simplify the
analysis, however, we will focus in this work on regimes where the
nonlinear mean-field interaction is negligible (cf.~\cite{AK,AK2}).
Effects of the mean-field interaction in related problems are
investigated in \cite{nonlinear_2d_PRL,nonlinear_2d_PRA}.

In order to conceive a 2D atom laser, we shall consider a condensate
loaded into a uniformly accelerated 2D optical square lattice in the
regime where Landau-Zener tunnelling is the predominant effect. In
particular, since the atoms need to tunnel out of the lattice to
form the lasing pulses, dynamics of the condensate at higher energy
bands will be essential (cf.~\cite{KK,germany04}). As we will see
below the more complicated energy landscape at higher energy bands
turns out to enrich the structure for the output pulses of the atom
laser. We shall study the dynamics of the condensate by numerically
solving the Gross-Pitaevskii (GP) equation (see e.g. \cite{PS}). For
appropriate accelerations we shall find pulses of atom clouds
tunnelling out of the condensate, which manifest a patterned
structure. We will show that these are consequences of Bragg
reflection and Landau-Zener tunnelling at higher energy bands. For
energy states in these higher energy bands, unlike their 1D
counterparts \cite{AK}, characteristics of Bloch waves still
manifests in the atom waves; even though they are quite close to
free-particle states. By means of $\bfk$-space analysis, we shall
relate propagation of the coherent atom waves in real space to the
$\bfk$-space dynamics of the condensate. As we will soon recognize,
aside from a 2D atom laser with patterned pulses, our result can
help envisage beam splitters for coherent atom waves.

In the following, we shall start in \sref{TF} with an introduction
to our theoretical formulation. Numerical results for our
simulations will then be presented in \sref{AL}. In \sref{kd} we
will try to understand these results base on $\bfk$-space dynamics
of the condensate. We will then demonstrate in \sref{manipln} how
this analysis can help provide a useful tool for manipulation of
coherent atom waves. A brief conclusion is supplied at the end of
the article.

\section{Two-dimensional atom laser}
\label{2dAL}
%=================================================
\subsection{Theoretical formulation}
\label{TF}
%=================================================
Let us consider a Bose-Einstein condensate of atoms with atomic mass
$m$ in a two dimensional optical lattice \cite{Greiner}. We shall
look into the regime where the low temperature ground-state dynamics
of the condensate can be described by the GP equation (see
e.g.~\cite{PS})
\begin{eqnarray}
i\hbar\frac{\partial}{\partial t} \Psi(\brho,t) =
\left( -\frac{\hbar^2}{2m} \nabla_\brho^2 + V_{ext}(\brho,t) +
U |\Psi(\brho,t)|^2 \right) \Psi(\brho,t)
\,,  \label{GP0}
\end{eqnarray}
where $\Psi$ is the condensate wave function, $\brho=(x,y)$ is the
2D position vector, $\nabla_\brho^2 \equiv \left(
\frac{\partial^2}{\partial x^2} + \frac{\partial^2}{\partial y^2}
\right)$ is the 2D Laplacian operator, $V_{ext}$ is the total
external potential acting on the condensate, and $U$ is the
mean-field coupling constant. In order to examine dynamics of the
condensate in an accelerated optical lattice, here the external
potential $V_{ext}$ consists of two parts: one due to the optical
lattice, and the other from a uniform acceleration. The optical
lattice potential that we shall consider is similar to that in
\cite{Greiner}
\begin{eqnarray}
V_{opt}(x,y) = V_0 \left[ \cos^2(q_x x) + \cos^2(q_y y) \right] \, ,
\label{Vopt}
\end{eqnarray}
where $V_0$ is a real constant, $q_x$, $q_y$ are wave numbers for
the laser beams that build up the optical lattice. In experiments
the lattice beams usually have the same wavelength, so that
$q_x=q_y=q$. We shall thus consider this case, so that \eref{Vopt}
corresponds to a 2D square lattice with lattice constant $a=\pi/q$.
If the lattice is uniformly accelerated with acceleration $\bfg$,
there would be the corresponding potential
\begin{eqnarray}
V_g(x,y) = - m \bfg\!\cdot\!\brho \, .
\label{Vg}
\end{eqnarray}

In order to maintain the phase coherence of the atom laser that we
have in mind, it is crucial to avoid large potential strengths
\cite{AK2}. This is because for high lattice strength $V_0$, if one
regards the condensate as a collection of individual atom clouds
over the lattice sites, with increasing $V_0$ the wave function
overlap between neighbouring sites would drop and, at the same time,
the mean-field interaction $U|\Psi|^2$ would grow due to the more
localized wave function and stronger coupling $U$ \cite{Jaksch}.
These effects combine to reduce the phase coherence among the
condensates on different lattice sites \cite{AK2,Orzel}. This can be
understood as a consequence of the number-squeezing of the
condensate on each lattice site due to the mean-field interaction
when $V_0$ is large, which enhances the (uncorrelated) phase
fluctuations among the lattice sites \cite{Orzel}. It is essential
to recognize the role of the mean-field interaction in this
decoherence mechanism: with $U=0$, as long as the wave function
overlap between adjacent sites remains finite, there would not be
any decoherence even for large $V_0$. We will thus be interested in
the regime where the mean-field interaction $U|\Psi|^2$ in
\eref{GP0} is negligible compared with the strength of the optical
potential.\footnote{Since the mean-field interaction is proportional
to the ($s$-wave) scattering length and the density of the
condensate, besides tuning the lattice strength $V_0$ (such as in
\cite{AK,AK2,Orzel,MST}), experimentally one could also attain this
regime by using condensates with low densities and/or reducing the
scattering length via Feshbach resonances \cite{FR1,FR2}.} In
experiments, this would correspond to the cases where well-resolved
Bragg peaks can be visible when the condensate is released from the
optical lattice \cite{Orzel,MST} (notice, however, the caveat
pointed out by \cite{interf}); namely the system is supposed to be
deep on the ``superfluid" side of the phase diagram. This would at
the same time also justify our analysis of the condensate dynamics
based on the GP equation, since for large potential strengths the
system could make transitions into the Mott insulating phase
\cite{MST}. Throughout this work, we will therefore consider the
following GP equation
\begin{eqnarray}
i\hbar\frac{\partial\Psi}{\partial t} = \left(
-\frac{\hbar^2}{2m} \nabla_\brho^2 + V_0 \left[ \cos^2(q x) + \cos^2(q y)
\right] - m \bfg\!\cdot\!\brho \right) \Psi \,. \label{GP}
\end{eqnarray}
The time evolution of the condensate can then be analyzed reliably
using the Crank-Nicholson method \cite{CN}, which we shall now turn
to.

\subsection{Atom laser}
\label{AL}
%=========================================

\begin{figure}
\begin{center}
\includegraphics*[width=160mm]{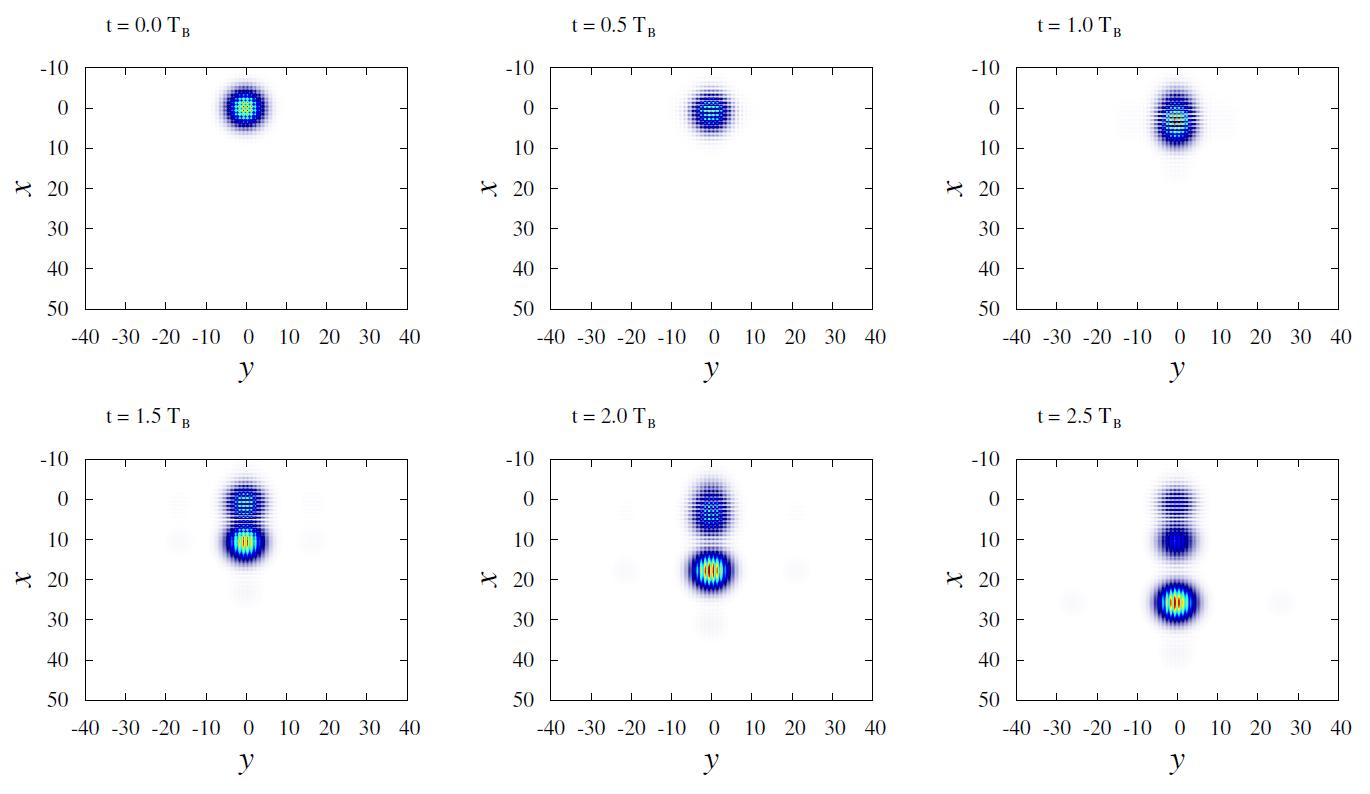}
\\ \hspace*{7mm} (a) \\
\includegraphics*[width=160mm]{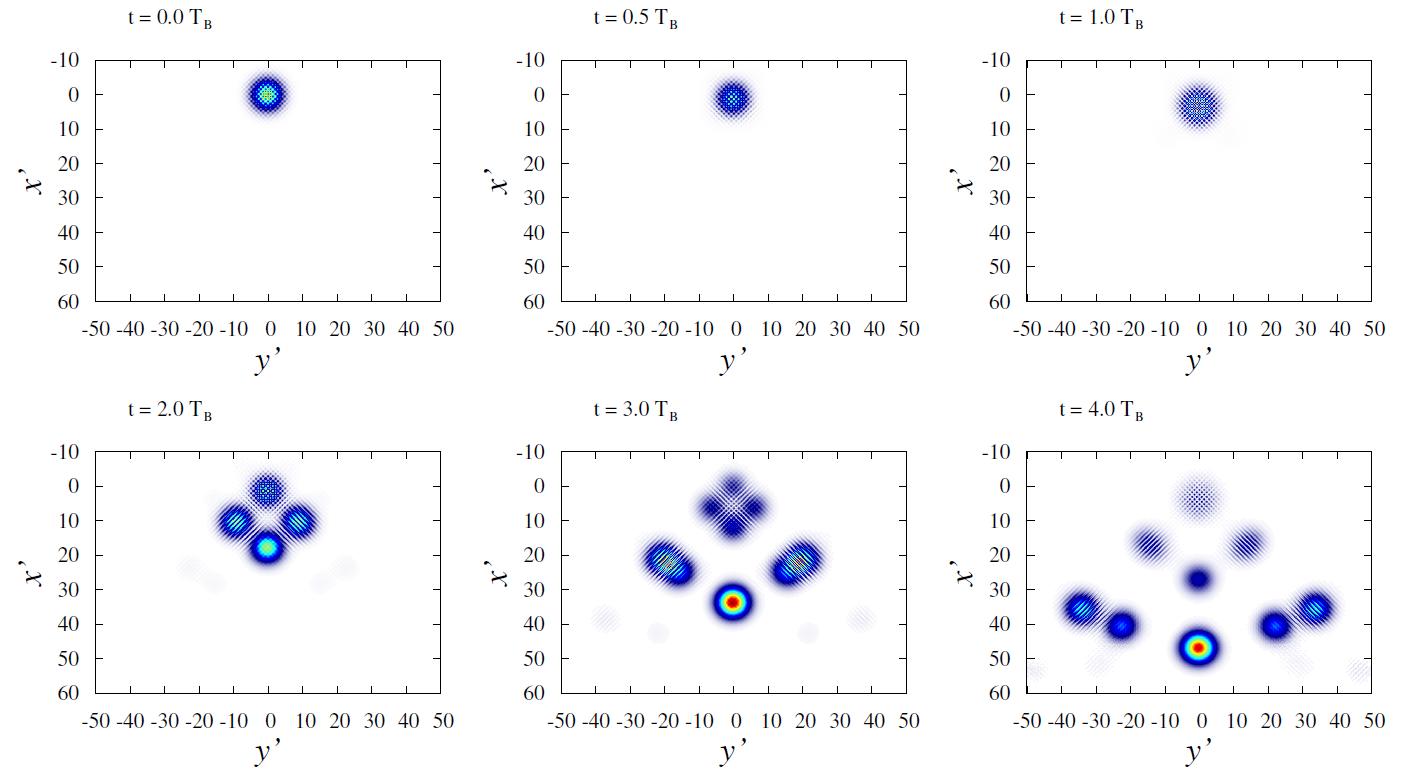}
\\ \hspace*{7mm} (b) \\
\end{center}
\caption{\small Numerical results for accelerated condensates in a 2D square lattice with the
acceleration $\bfg$ along
(a) the $(1,0)$ direction and (b) the (1,1) direction. For both cases we have $V_0 = 1.0 E_R$
and $g=\frac{\pi}{4} \frac{E_R}{ma}$, and the time scale $T_B$ is given by \eref{Tj}. Notice that
in (b), for convenience, the plots are made with respect to the coordinates $x'=(x+y)/\sqrt{2}$,
$y'=(y-x)/\sqrt{2}$, so that $\bfg$ points along the $x'$-axis. In all plots the coordinates are
measured in units of the lattice constant $a$.
\label{2dAL_fig} }
\end{figure}

Let us suppose we have initially in free space a stationary
condensate which has a Gaussian wave function
\begin{eqnarray}
\Psi(\rho,t_0) = \sqrt{\frac{N}{2\pi\sigma^2}}\exp\!\left(-\frac{\rho^2}{4\sigma^2}\right) \, ,
\label{gaussian_0}
\end{eqnarray}
where $N$ is the total number of atoms, $\rho=(x^2+y^2)^{1/2}$, and
$\sigma$ is the width of the wave packet. In order to transfer the
condensate into the lattice ground state, without turning on $V_g$,
the Gaussian wave packet is first subject to the optical lattice
potential $V_{opt}$ and set to evolve for a period of time.  At time
$t=0$, we switch on $V_g$ [see \eref{Vg}] so that a static uniform
acceleration $\bfg$ starts acting on the condensate. It is then our
main purpose to examine the subsequent dynamics of the condensate.
As explicated previously, to avoid complications from strong lattice
potentials we restrict here to potential strengths of order of the
recoil energy
\begin{eqnarray}
E_R \equiv \frac{\hbar^2 q^2}{2m} = \frac{h^2}{8m a^2} \, .
\end{eqnarray}

We carry out numerical simulations for the above procedures base on
the GP equation \eref{GP}. The results for two different
accelerations are presented in figure \ref{2dAL_fig}, where we have
used $\sigma=2.5 a$ and $N= 2\pi\sigma^2$ for the initial wave
function \eref{gaussian_0}, and $V_0 = 1.0 E_R$ for the optical
potential \eref{Vopt}. The acceleration is set to have the magnitude
$g=\frac{\pi}{4} \frac{E_R}{ma}$, which has been chosen somewhat
arbitrarily; one only has to make sure that it would lead to
appreciable tunnelling probability for the condensate (see later).
For convenience, we use the following time scale for the time
evolution of the condensate
\begin{eqnarray}
T_B = \frac{\left(\frac{2\pi}{a}\right)}{\left(\frac{mg}{\hbar}\right)}
= \frac{h}{mga}
= \frac{1}{g} \sqrt{\frac{8E_R}{m}} \, ,
\label{Tj}
\end{eqnarray}
which is the time for a Bloch state to cross one complete Brillouin
zone (namely, the Josephson time \cite{AK}, or the Bloch period
\cite{germany04}) when $\bfg$ is along the lattice $(1,0)$
direction.\footnote{Notice that for $\bfg$ along arbitrary
directions, $T_B$ would be in general not identical to the period
for the Bloch oscillation. In fact, it is possible to have open
trajectories for Bloch oscillations \cite{germany04}. In this case,
the Bloch period does not exist.} We shall henceforth adopt a
coordinate system with $x$-axis along the lattice $(1,0)$ direction,
and $y$-axis the $(0,1)$ direction.

As shown in figure \ref{2dAL_fig}(a), for $\bfg$ along the $(1,0)$
direction pulses of atom clouds tunnel out of the condensate in a
manner reminiscent of that in accelerated 1D optical lattices
\cite{AK,AK2}. The sequence of pulses generated are equally spaced
in time by $T_B$ and aligned entirely in the $(1,0)$ direction. For
$\bfg$ along the $(1,1)$ direction [see figure \ref{2dAL_fig}(b)],
however, the pulses have a more complicated structure. There are
initially three pulses generated from the condensate in three
different directions which subsequently split into further
sub-pulses. Clearly, this immediately brings up interesting
possibilities for applications: besides a ``patterned" atom laser,
one could also use the 2D optical lattice as a beam splitter for
coherent atom waves, as we will soon demonstrate. In the following
we shall first examine the origin of this pattern and afterwards
elaborate on its possible applications.

\subsection{$\bfk$-space dynamics}
\label{kd}
%==================================
We shall now attempt to understand the results above from the
viewpoint of $\bfk$-space dynamics of the condensate. This analysis
will turn out very useful for real-space manipulations of atom waves
using accelerated 2D optical lattices.

In the absence of lattice acceleration (i.e. $g=0$), the eigenstates
of the GP equation \eref{GP} for the condensate are the Bloch states
(see e.g.~\cite{AM}). Each Bloch state is characterized by the
quantum numbers $\bfk$ (the Bloch wave-vector) and $n$ (the band
index). For the initial wave function \eref{gaussian_0}, it
corresponds in $\bfk$-space to a Gaussian centering at $\bfk=(0,0)$
with a spread $\delta k_x = \delta k_y = (2\sigma)^{-1}$. Therefore,
the subsequent time evolution of the condensate would yield wave
packets that are superpositions of Bloch states with width $\delta k
\sim (2\sigma)^{-1}$. For the simulations in figure \ref{2dAL_fig}
we have $\sigma=2.5 a$, thus in $\bfk$-space, instead of a single
point, each atom cloud corresponds to a wave packet that spreads
around a central value with area $\pi (\delta k)^2 \sim
\frac{\pi}{4\sigma^2}=\frac{\pi}{25 a^2}$, which is
$\frac{1}{100\pi}\simeq 0.32\%$ the area of one single Brillouin
zone. In the analysis below, we shall ignore the spread in $\bfk$
and focus on the center of each wave packet, as if the atom cloud
were in a sharp Bloch state characterized by a single Bloch wave
vector. In a more quantitative treatment, one can take into account
the spreading by averaging over the distribution of the Bloch wave
vectors \cite{Dahan}.

In view of the above, the motion of the condensate under the action
of a uniform static acceleration $\bfg$ can then be dealt with using
a semiclassical approach. Within each band (namely, for one specific
$n$) the dynamics of the condensate follows the equations of motion
(see e.g.~\cite{AM})
\begin{eqnarray}
\frac{d {\brho}}{d t} &=& \frac{1}{\hbar} \nabla_\bfk E_n \, ,
\label{semiclassical_1}
\\
\hbar\frac{d \bfk}{d t}  &=& m {\boldsymbol g} \, ,
\label{semiclassical_2}
\end{eqnarray}
where $\brho$ is, as before, the 2D position vector, and $E_n$ is
the eigenenergy for the Bloch state. As is clear from
\eref{semiclassical_2}, under the action of a constant acceleration
$\bfg$ the Bloch state would proceed steadily in $\bfk$-space along
the direction determined by $\bfg$. However, when the state arrives
at zone boundaries where energy gaps exist, the state would have to
either make transitions to higher energy bands (Landau-Zener
tunnelling) or be reflected to other regions of the same zone (Bragg
reflection), so that it could continue its course of evolution
according to \eref{semiclassical_2}. Notice that these processes are
not included in \eref{semiclassical_2} since two or more energy
bands can be involved. When only two states are involved, one can
find that the tunnelling probability would be (see e.g.~\cite{Ziman})
\begin{eqnarray}
P = \exp\left(-\frac{\pi^2}{4} \frac{E_{gap}^2}{mga E_0}\right) \, ,
\label{LZ}
\end{eqnarray}
where $E_{gap}$ is the energy gap at the $\bfk$-point in question,
and $E_0$ is the corresponding free-particle energy. If the
acceleration $g$ is small, the tunnelling probability would be
exponentially small. The state would then be Bragg reflected to
another state within the original zone, and subsequently continue to
evolve in accordance with \eref{semiclassical_1},
\eref{semiclassical_2} until it meets another zone boundary. If the
state has low probability for interband transitions, its motion
would be limited to a single zone and the motion would be
oscillatory in both real and $\bfk$ spaces (which is the Bloch
oscillation) \cite{germany04}. However, for larger values of $g$ the
tunnelling probability can become appreciable. The state can then
tunnel into higher energy bands and give rise to pulses of atom
clouds in real space moving with group velocities determined by
\eref{semiclassical_1}. As we will find out soon, the patterns of
the atom pulses in figure \ref{2dAL_fig} are consequences of Bragg
reflection and Landau-Zener tunnelling of the atom waves. Base on
this observation we will explain below, how this picture can be used
to carry out coherent manipulations of atom waves.

\begin{figure}
\begin{center}
\hspace*{10mm}
\includegraphics*[width=70mm]{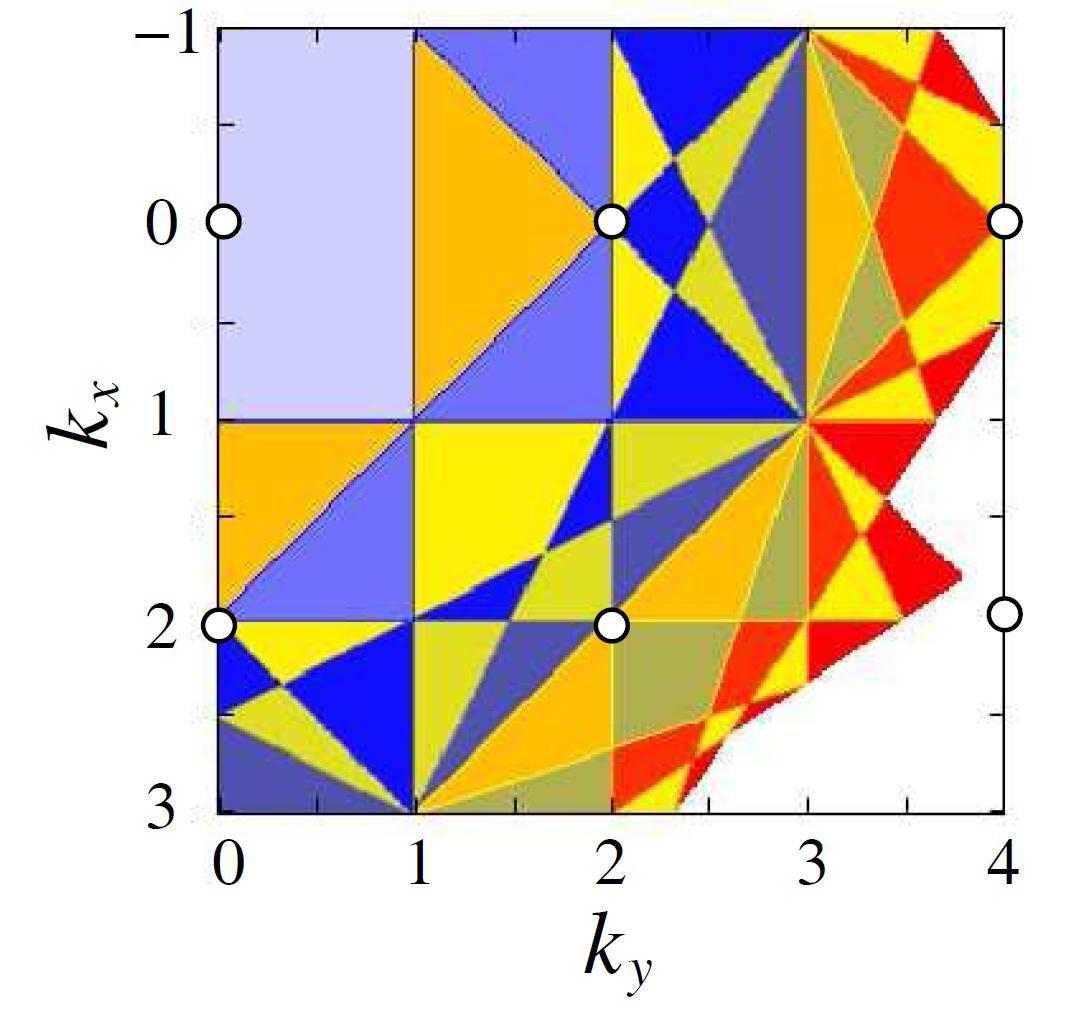}
\hspace*{5mm}
\includegraphics*[width=65mm]{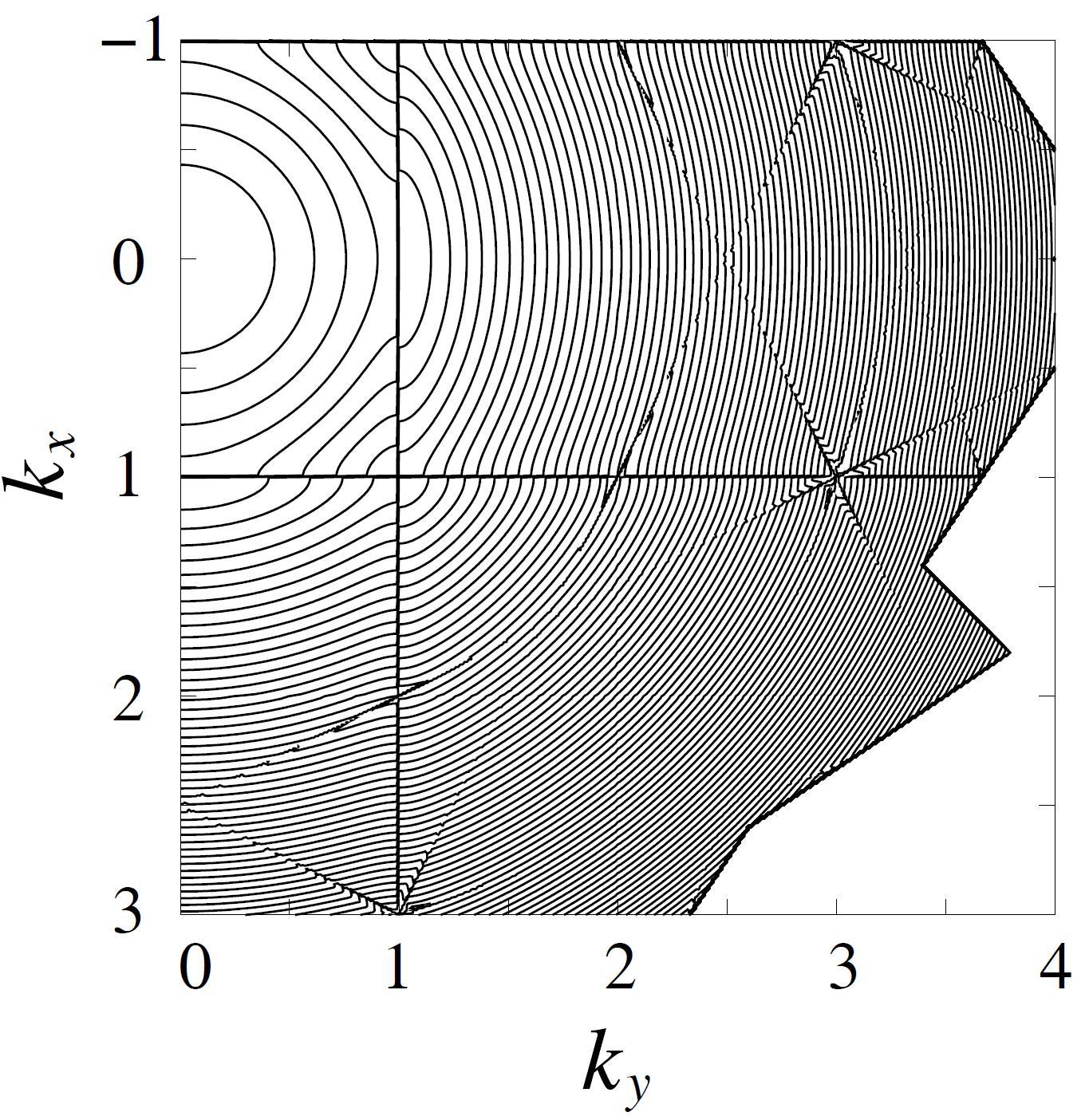}
\\ \hspace*{21mm} (a) \hspace{66mm} (b)
\end{center}
\caption{\small (a) The first twelve Brillouin zones for the 2D square lattice in the extended-zone scheme.
The open circles represent reciprocal lattice points. (b) shows contours for the corresponding energy
landscape when $V_0=1.0 E_R$. Here $k_x$, $k_y$ are both in units of $\pi/a$. For regions not shown here,
one can infer the zone structure and energy landscape up to the twelfth zone from symmetry of the lattice.
\label{2dBZ} }
\end{figure}

On the grounds of the foregoing analysis, we note that the state of
the condensate migrates over the $\bfk$-space under the pull of
$\bfg$ and hops around via Bragg reflections. It is thus very
helpful to examine the $\bfk$-space dynamics of the condensate using
the extended-zone picture (see e.g.~\cite{Ziman}). Figure~\ref{2dBZ}
shows the extended-zone energy landscape over $\bfk$-space obtained
from direct diagonalization for the condensate in the 2D square
lattice \eref{Vopt}. Energy gaps exist at regions where there are
discontinuities in the energy contours, which appear as dark shading
lines in figure \ref{2dBZ}(b); the darker the shading is, the larger
the energy gap is. Notice that not every zone boundary has energy
gaps. As in the extended-zone scheme the zone index is identical to
the band index, we shall henceforth use them interchangeably.

In our simulations in figure \ref{2dAL_fig}, we have chosen
$g=\frac{\pi}{4}\frac{E_R}{ma}$, which is sufficiently large to
ensure that the tunnelling probability out of the first band is
always appreciable for $V_0=1.0 E_R$. With the initial wave function
\eref{gaussian_0}, the condensate starts at $t=0$ from the state
$\bfk=(0,0)$ in the first zone. For figure \ref{2dAL_fig}(a) the
acceleration points along the $(1,0)$ direction, therefore according
to \eref{semiclassical_2} the condensate would proceed along the
$k_x$-axis until it meets the zone boundary at $k_x=\frac{\pi}{a}$.
The wave packet then becomes partly reflected back to the first zone
and partly transmitted into the second zone, giving rise to the
first tunnelling pulse shown in figure \ref{2dAL_fig}(a). The pulse
that enters the second zone proceeds again according to
\eref{semiclassical_2} until it meets the zone boundary at
$\bfk=(\frac{2\pi}{a},0)$, where three Bragg planes intersect. From
figure \ref{2dBZ}(b) it is clear that the energy gap at this point
is tiny, thus the pulse can tunnel through entirely. As the wave
packet moves along, its energy also grows. Eventually the wave
packet becomes essentially free, since it can overcome any energy
gap that it encounters [see \eref{LZ}]. Notice that this result is
essentially identical to that for 1D optical lattices \cite{AK,AK2}.

For acceleration along the $(1,1)$ direction [see figure
\ref{2dAL_fig}(b)], the $\bfk$-space dynamics for the condensate
becomes more complicated. Again, the condensate starts at $t=0$ from
the state $\bfk=(0,0)$ and evolves initially according to
\eref{semiclassical_2} along the $(1,1)$ direction. However, at
$t=T_B/\sqrt{2}$ the state reaches the point
$\bfk=(\frac{\pi}{a},\frac{\pi}{a})$, where bands $1-4$ meet. Due to
the large energy gap, the condensate can have a partial wave Bragg
reflected back into the first band [to the point
$\bfk=(-\frac{\pi}{a},-\frac{\pi}{a})$] and a partial wave
tunnelling into the second band [or the third band, since they are
degenerate at this point; see figure \ref{2dBZ}(b)]. For the partial
wave entering into the second band, nevertheless, since the
acceleration $\bfg$ would drive the system to evolve along the
$(1,1)$ direction, the state has to either tunnel immediately into
the fourth band, or be Bragg reflected within band 2. The four atom
clouds visible at $t=2.0\,T_B$ in figure \ref{2dAL_fig}(b) represent
each of the above possibilities: the atom cloud at the top corner is
the partial wave that is Bragg reflected back into the first band,
executing an oscillatory motion in both real and $\bfk$-spaces; the
lower one is the partial wave that tunnels directly into the fourth
band. The two side pulses are partial waves that are Bragg reflected
to the points $\bfk=(\pm\frac{\pi}{a},\mp\frac{\pi}{a})$ which
subsequently evolve under the acceleration $\bfg$ towards the states
$(\frac{2\pi}{a},0)$ and $(0,\frac{2\pi}{a})$, respectively. For the
three partial waves that tunnel out of the first band, the
acceleration $\bfg$ continues to bring them forward along the
$(1,1)$ direction. The central pulse becomes essentially free, as
there is no longer any appreciable energy gap along its $\bfk$-space
trajectory. For the two side pulses, however, as one can see from
figure \ref{2dBZ}(b) there remain several gaps that they shall
cross.

Since the two side pulses are symmetric, it will suffice to focus
our analysis on one of them. Let us consider the one on the right.
After being Bragg reflected to the point
$(-\frac{\pi}{a},\frac{\pi}{a})$, the pulse evolves under the pull
of $\bfg$ along the $(1,1)$ direction. It will therefore come across
an energy gap at $\bfk=(\frac{\pi}{3a},\frac{7\pi}{3a})$ at the time
$t=\frac{7\sqrt{2}}{6}\,T_B\simeq 1.650\,T_B$. For the result in
figure \ref{2dAL_fig}(b), one can rule out any possibility for Bragg
reflection at this point since the sub-pulses become visible only at
$t=2.0\sim 3.0\,T_B$ (for a more detailed check, see later). The
next energy gap that the pulse would meet occurs at
$t=\frac{3}{\sqrt{2}}\,T_B\simeq 2.121\,T_B$ at the point
$\bfk=(\frac{\pi}{a},\frac{3\pi}{a})$. As we shall explain in
greater detail below, at this point the pulse turns out to be
partially Bragg reflected to the state
$\bfk=(-\frac{\pi}{a},\frac{3\pi}{a})$ and partially tunnels forward
along its original trajectory. Therefore, the original pulse is
split into two sub-pulses, resulting in the final pattern seen in
figure \ref{2dAL_fig}(b).

To substantiate the arguments in the preceding paragraph, one could
consider an arbitrary Bloch state on the line segment between the
$\bfk$ points $(-\frac{\pi}{a},\frac{\pi}{a})$ and
$(\frac{\pi}{3a},\frac{7\pi}{3a})$, and apply to the state the same
acceleration $\bfg$ as above, but now in a {\em controlled} manner.
For instance, starting from the point
$\bfk=(-\frac{\pi}{2a},\frac{3\pi}{2a})$, we apply the acceleration
$\bfg$ for the duration $\Delta t=1.0\,T_B$ and then turn it off.
This would send the pulse to the state $\bfk =
(\frac{-1+2\sqrt{2}}{2}\,\frac{\pi}{a},\frac{3+2\sqrt{2}}{2}\,\frac{\pi}{a})
\simeq (\frac{0.914\pi}{a},\frac{2.914\pi}{a})$, which lies between
the points $(\frac{\pi}{3a},\frac{7\pi}{3a})$ and
$(\frac{\pi}{a},\frac{3\pi}{a})$, and then leave it to evolve freely
under the lattice potential \eref{Vopt}. We find that in this case
the atom cloud remains a single one throughout its time evolution.
If now the acceleration is kept on for a longer period of time, so
that the pulse could go beyond the point
$\bfk=(\frac{\pi}{a},\frac{3\pi}{a})$ slightly, one would observe
that the two sub-pulses are generated. To confirm that Bragg
reflection and Landau-Zener tunnelling do occur at
$(\frac{\pi}{a},\frac{3\pi}{a})$, one could repeat the above
procedures but apply now a reduced acceleration along $(1,1)$
direction, say, $g'=\frac{\pi}{20}\frac{E_R}{ma}$. One would find in
this case only one single pulse appears, even though an acceleration
duration that could have sent the state beyond $\bfk =
(\frac{\pi}{a},\frac{3\pi}{a})$ had been applied. What happens here
is that the pulse is Bragg reflected from
$(\frac{\pi}{a},\frac{3\pi}{a})$ to the state
$(-\frac{\pi}{a},\frac{3\pi}{a})$, and then tunnels into the next
band under the action of $\bfg'$.\footnote{There are in fact other
possible final states for the Bragg reflection \cite{AM}. However,
they are suppressed here since the optical potential \eref{Vopt} has
Fourier components only for $\Delta \bfk=(0,0)$, $(\pm 2\pi/a,0)$,
and $(0,\pm 2\pi/a)$.} To further endorse our argument, we notice
that the directions of the group velocities of the wave packets,
according to \eref{semiclassical_1}, can be read off from the
contours in figure \ref{2dBZ}(b). Hence the sub-pulse that tunnels
through from $\bfk = (-\frac{\pi}{a},\frac{3\pi}{a})$ has a group
velocity almost parallel to the $(0,1)$ direction, while the one
that tunnels forward from $\bfk = (\frac{\pi}{a},\frac{3\pi}{a})$
would be more free-particle like [i.e.~its real space trajectory
would bend downwards just like a projectile; see figure
\ref{2dAL_fig}(b)]. From the perspective of energetics, we notice
that along the $(1,1)$ direction, tunnelling across the point
$(-\frac{\pi}{a},\frac{3\pi}{a})$ would send the wave packet into
the ninth band (degenerate with the tenth band), while across
$(\frac{\pi}{a},\frac{3\pi}{a})$ would bring it into the eleventh
band (degenerate with the twelfth band). Taking into account the
finite width of the wave packet, one can find that the former option
is evidently energetically more favourable. It is also possible to
confirm our argument base on Fourier analysis of the atom clouds
\cite{THW}.

Although a more quantitative analysis of the problem than that
presented above is possible \cite{Shchesnovich}, our approach in the
foregoing paragraph has its merits. Besides providing a check for
our argument it also alludes to one important point: in addition to
considering condensates with zero (center-of-mass) initial velocity
(as was done for the simulations in figure \ref{2dAL_fig}), one
could also consider an initial wave function with non-zero wave
vector. Our analysis thus brings out a useful means for the
manipulation of coherent atom waves, as we shall now explain.

\section{Manipulation of coherent atom waves}
\label{manipln}
%=================================================
In order to prepare a condensate with non-zero Bloch wave vector
$\bfk_0$, one can start from a stationary wave packet with the wave
function \eref{gaussian_0}. Without turning on the optical lattice
potential, one applies first an acceleration $\bfg_0$ (along the
$\bfk_0$ direction) to the stationary condensate for the period of
time $\Delta t=\frac{\hbar k_0}{mg_0}$ and then switch it off. The
condensate would now have acquired the designated wave vector
$\bfk_0$ and the wave function would be
\begin{eqnarray}
\Psi(\brho) = \sqrt{\frac{N}{2\pi\sigma^2}}\exp\!\left(-\frac{\rho^2}{4\sigma^2}\right)
\exp\!\left(i \bfk_0\cdot\brho \right)\, .
\label{gaussian_v}
\end{eqnarray}
To project the state onto the lattice eigenstate with Bloch wave
vector $\bfk_0$, one could switch on the optical lattice and allow
the state \eref{gaussian_v} to evolve for a period of time. This
would then transfer the condensate into the desired Bloch state. It
should be noticed that this procedure for state preparation would
work better for weak lattice potentials (and $\bfk_0$ a distance
away from any energy gaps, of course). For strong lattice potentials
the Gaussian state \eref{gaussian_v} may deviate enormously from the
Bloch state $\bfk_0$; thus the projection procedure may result in an
atom cloud with extremely low density. To overcome this difficulty,
one may first project the state onto a weak-potential eigenstate and
afterwards increase the lattice strength steadily.

\begin{figure}
\begin{center}
\includegraphics*[width=160mm]{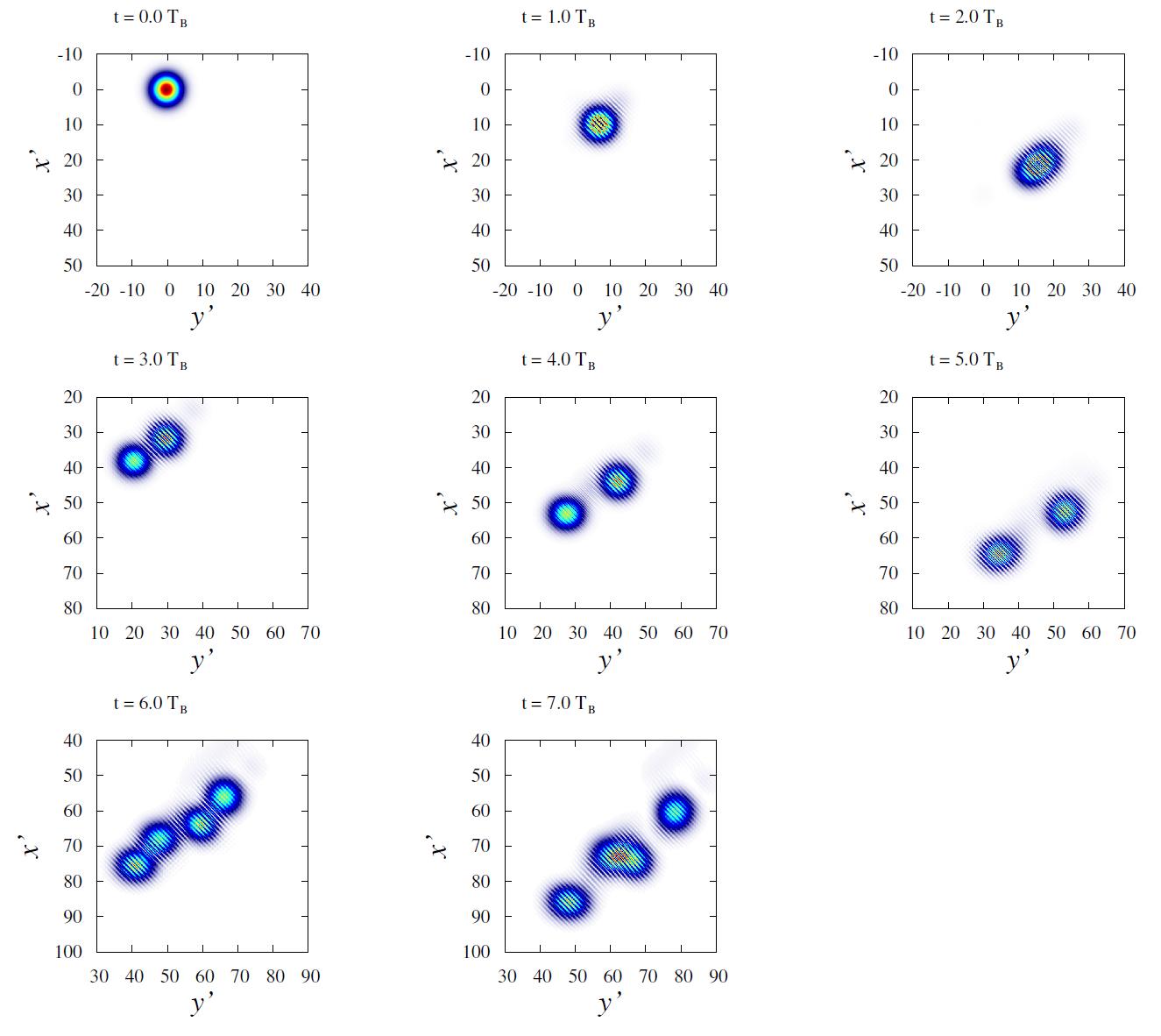}
\\*[-50mm]
\end{center}
\caption{\small Numerical simulation for an accelerated condensate
that starts from the initial wave function \eref{gaussian_v} with
$\bfk_0=(\frac{\pi}{2a},\frac{5\pi}{2a})$ in the same 2D optical
lattice as in figure \ref{2dAL_fig}. An acceleration
$g=\frac{\pi}{4}\frac{E_R}{ma}$ is applied along the $(1,1)$
direction in a controlled manner as described in the text. As in
figure \ref{2dAL_fig} (b), here the time scale $T_B$ is given by
\eref{Tj} and the plots are made against the rotated coordinates
$x'$ and $y'$ in units of the lattice constant $a$. Notice that for
different time frames we have shifted the coordinates to different
ranges for convenience. \label{4_beam} }
\end{figure}

In figure \ref{4_beam} we demonstrate manipulation of coherent atom
waves utilizing the scheme proposed in the previous section. Here we
start at $t=0$ from the Gaussian wave packet \eref{gaussian_v} with
$\sigma=2.5 a$, $N= 2\pi\sigma^2$, and
$\bfk_0=(\frac{\pi}{2a},\frac{5\pi}{2a})$. The state is left to
evolve in the optical lattice \eref{Vopt} with $V_0=1.0 E_R$ for the
period of time $\Delta t= 1.0\,T_B$. We then turn on the
acceleration $g=\frac{\pi}{4}\frac{E_R}{ma}$ along the $(1,1)$
direction for the duration $\Delta t= 1.0\,T_B$, sending the state
to the point $\bfk =
(\frac{1+2\sqrt{2}}{2}\,\frac{\pi}{a},\frac{5+2\sqrt{2}}{2}\,\frac{\pi}{a})
\simeq (\frac{1.914\pi}{a},\frac{3.914\pi}{a})$, and then switch it
off. The condensate is then left to evolve in the lattice potential
for another period of time $\Delta t= 2.0\,T_B$. As one can see in
figure \ref{4_beam}, the condensate begins to split at $t=2.0\,T_B$
and two pulses of atom clouds become visible in the plot for
$t=3.0\,T_B$. To further split the two pulses, at $t=4.0\,T_B$ we
apply the acceleration $-\bfg$, which is opposite to that applied
earlier. We turn off the acceleration after the duration $\Delta t=
1.0\,T_B$ and allow the pulses to evolve in the lattice freely. The
reverted acceleration brings the two pulses across the
beam-splitting points $\bfk = (\pm\frac{\pi}{a},\frac{3\pi}{a})$.
Therefore, as shown in figure \ref{4_beam}, at $t=6.0\,T_B$ each of
the two pulses splits again into two further sub-pulses. All these
results are consequences of the processes expounded in the previous
section. One can see that by exploring Bragg reflection and
Landau-Zener tunnelling of the condensate in accelerated 2D optical
lattices, we have furnished a beam splitter that may be useful for
manipulation of atom waves. By changing the lattice strength $V_0$
and/or the magnitude of the acceleration $g$, one can tune the
relative strengths between the reflected and the transmitted pulses.
Further manipulations of the atom cloud can be achieved by extensive
exploration of the energy landscape for the optical lattice. For
instance, one can apply $\bfg$ in arbitrary directions, change
$\bfg$ in real time, and/or make use of different lattice structures
(which may also involve ``defects"). A simple example for the
interference of coherent atom waves is shown in the plot for
$t=7.0\,T_B$ in figure \ref{4_beam}.

\section{Conclusions and discussions}
\label{fin}
%==============================
The interplay between cold atom physics and solid-state physics has
stimulated exciting developments in the past decades. In this work,
we have studied the dynamics of a cloud of ultracold atoms subject
to an accelerated 2D optical square lattice by numerically solving
the GP equation in the regime where mean-field interaction is
negligible. We have shown that for sufficiently large accelerations
the condensate can generate pulses of atom clouds that have
structures richer than its 1D counterparts. Using a semiclassical
picture, we have demonstrated that these structures can be
understood from the $\bfk$-space dynamics of the Bloch state. In
particular, Bragg reflection and Landau-Zener tunnelling at higher
energy bands are shown to be responsible for the pattern generated.

On the one hand, our result can help envisage 2D atom lasers that
have patterned output pulses. These patterns are closely connected
with the underlying lattice structures. On the other hand, we have
shown that by appropriately controlling the magnitude, direction,
and duration of the lattice acceleration, one would be able to
manipulate the atom wave by exploring the energy landscape of the
optical lattice. We have demonstrated that a beam splitter for
coherent atom waves can be furnished in this scheme.

An important issue that we did not address in this article is the
effect of the nonlinear mean-field interaction in \eref{GP0}.
Besides decoherence of the condensate as discussed in section 2.1
\cite{AK2,Orzel}, it has been shown that nonlinear effects can cause
asymmetric Landau-Zener tunnelling and modulation instability for
condensates in accelerated 2D optical lattices
\cite{nonlinear_2d_PRL,nonlinear_2d_PRA}. We hope to investigate
these problems in future publications. At the same time, possible
applications ensue from this work in the fields of atom optics and
quantum information sciences are also yet to be explored.

%--------------------------------------------------------------------
\ack The authors would like to thank Profs. Sungkit Yip , Chung-Yu
Mou, and Chien-Hua Pao for valuable discussions. This research was
supported by NSC of Taiwan through grant numbers NSC
96-2112-M-194-011-MY3, and NSC 98-2112-M-194-001-MY3; it is also
partly supported by the Center for Theoretical Sciences, Taiwan.

%========================= [REFERENCES] ==============================


\section*{References}
\begin{thebibliography}{99}
\bibitem{bloch}
Bloch I 2008 {\em Science} {\bf 319} 1202

\bibitem{MO}
Morsch O and Oberthaler M 2006
{\em Rev. Mod. Phy.} {\bf 78} 179

\bibitem{AM}
Ashcroft N W and Mermin N D 1976
{\em Solid State Physics} (Singapore: Thomson Learning)

\bibitem{Ziman}
Ziman J M 1972
{\em Principles of the Theory of Solids} 2nd ed (Cambridge: Cambridge University Press)

\bibitem{RSQ}
Raizen M, Salomon C and Niu Q 1997
{\em Phys. Today} {\bf 50}(7) 30

\bibitem{PS}
Pitaevskii L and Stringari S 2003 {\em Bose-Einstein Condensation}
(Oxford: Oxford University Press)

\bibitem{Couvert}
Couvert A, Jeppesen M, Kawalec T, Reinaudi G, Mathevet R and Gu\'{e}ry-Odelin D 2008
{\em Europhys. Lett.} {\bf 83} 50001

\bibitem{Dahan}
Ben Dahan M, Peik E, Reichel J, Castin Y and Salomon C 1996
{\em Phys. Rev. Lett.} {\bf 76} 4508

\bibitem{WS}
Wilkinson S R, Bharucha C F, Madison K W, Niu Q and Raizen M G 1996
{\em Phys. Rev. Lett.} {\bf 76} 4512

\bibitem{LZ1d}
Bharucha C F, Madison K W, Morrow P R, Wilkinson S R, Sundaram B and Raizen M G 1997
{\em Phys. Rev.} A {\bf 55} R857

\bibitem{AK}
Anderson B P and Kasevich M A 1998 {\em Science} {\bf 282} 1686

\bibitem{AK2}
Anderson B P and Kasevich M A 1999 in
{\em Proceedings of the International School of Physics ``Enrico Fermi", Course CXL}
ed M Inguscio {\em et al} (Amsterdam: IOS Press) pp 439-452

\bibitem{KK}
Kolovsky A R and Korsch H J 2003
{\em Phys. Rev.} A {\bf 67} 063601

\bibitem{germany04}
Witthaut D, Keck F, Korsch H J and Mossmann S 2004
{\em New J. Phys.} {\bf 6} 41

\bibitem{germany07}
Breid B M, Witthaut D and Korsch H J 2007
{\em New J. Phys.} {\bf 9} 62

\bibitem{Trompeter}
Trompeter H, Krolikowski W, Neshev D N, Desyatnikov A S, Sukhorukov
A A, Kivshar Y S, Pertsch T, Peschel U and Lederer F 2006 {\em Phys.
Rev. Lett.} {\bf 96} 053903

\bibitem{Shchesnovich}
Shchesnovich V S, Cavalcanti S B, Hickmann J M and Kivshar Y S 2006
{\em Phys. Rev.} E {\bf 74} 056602

\bibitem{SDK}
Shchesnovich V S, Desyatnikov A S and Kivshar Y S 2008 {\em Opt.
Express} {\bf 16} 14076

\bibitem{Dreisow}
Dreisow F, Szameit A, Heinrich M, Pertsch T, Nolte S and
T\"{u}nnermann A 2009 {\em Phys. Rev. Lett.} {\bf 102} 076802

\bibitem{nonlinear_2d_PRL}
Brazhnyi V A, Konotop V V and Kuzmiak V 2006
{\em Phys. Rev. Lett.} {\bf 96} 150402

\bibitem{nonlinear_2d_PRA}
Brazhnyi V A, Konotop V V, Kuzmiak V and Shchesnovich V S 2007
{\em Phys. Rev.} A {\bf 76} 023608

\bibitem{Greiner}
Greiner M, Bloch I, Mandel O, H\"{a}nsch T W and Esslinger T 2001
{\em Phys. Rev. Lett.} {\bf 87} 160405

\bibitem{Jaksch}
Jaksch D, Bruder C, Cirac, J I, Gardiner C W and Zoller P 1998
{\em Phys. Rev. Lett.} {\bf 81} 3108

\bibitem{Orzel}
Orzel C, Tuchman A K, Fenselau M L, Yasuda M and M A Kasevich 2001
{\em Science} {\bf 291} 2386

\bibitem{MST}
Greiner M, Mandel O, Esslinger T, H\"{a}nsch T W and Bloch I 2002
{\em Nature} {\bf 415} 39

\bibitem{FR1}
Inouye S, Andrews M R, Stenger J, Miesner H-J, Stamper-Kurn D M and Ketterle W 1998
{\em Nature} {\bf 392} 151

\bibitem{FR2}
Chin C, Grimm R, Julienne P and Tiesinga E 2010
{\em Rev. Mod. Phys.} {\bf 82} 1225

\bibitem{interf}
Hadzibabic Z, Stock S, Battelier B, Bretin V, and Dalibard J 2004
{\em Phys. Rev. Lett.} {\bf 93} 180403

\bibitem{CN}
Press W H, Teukolsky S A, Vetterling W T and Flannery B P 2002
{\em Numerical Recipes in C++: The Art of Scientific Computing} 2nd ed
(Cambridge: Cambridge University Press)

\bibitem{THW}
Ting W C, Han D J and Wu S T 2010 Work in progress

\end{thebibliography}
\end{document}